\documentclass[twoside,onecolumn]{article}

\usepackage{blindtext} 

\usepackage[sc]{mathpazo} 
\usepackage[T1]{fontenc} 
\usepackage{microtype} 

\usepackage[english]{babel} 

\usepackage[hmarginratio=1:1,top=32mm,columnsep=20pt]{geometry} 
\usepackage[hang, small,labelfont=bf,up,textfont=it,up]{caption} 
\usepackage{booktabs} 

\usepackage{lettrine} 

\usepackage{enumitem} 
\setlist[itemize]{noitemsep} 

\usepackage{abstract} 

\usepackage{titlesec} 
\renewcommand\thesection{\Roman{section}} 
\renewcommand\thesubsection{\roman{subsection}} 
\titleformat{\section}[block]{\large\scshape\centering}{\thesection.}{1em}{} 
\titleformat{\subsection}[block]{\large}{\thesubsection.}{1em}{} 

\usepackage{fancyhdr} 
\usepackage{graphics}
\usepackage{graphicx}
\usepackage{epsfig}
\usepackage{amssymb}
\usepackage{cite}
\usepackage{titling} 

\usepackage{hyperref} 

\pretitle{\begin{center}\Huge\bfseries} 
\posttitle{\end{center}} 
\title{A comprehensive bifurcation method to analyze the super-harmonic and ultra-harmonic
behavior of the acoustically excited bubble oscillator} 
\author{%
\textsc{AJ. Sojahrood, D. Wegierak, H. Haghi, R. Karshafian and M.C. Kolios} \\[1ex]
\normalsize Department of Physics, Ryerson University, Toronto, Ontario, Canada \\ 
\normalsize \href{mailto:amin.jafarisojahrood@ryerson.ca}{amin.jafarisojahrood@ryerson.ca} 
}
\date{\today} 
\renewcommand{\maketitlehookd}{%
\begin{abstract}
Acoustically excited bubbles are involved in a wide range of phenomena and
applications ranging from oceanography to sonoluminescence; they have
applications in chemistry, medical imaging, and therapeutic ultrasound. The
complexity of bubble dynamics and the limited understanding of their behavior
restricts the exploration of their full potential. The bubble oscillator is a
highly nonlinear system, which makes it difficult to generate a comprehensive
understanding of its oscillatory behavior. One method used to investigate such
complex dynamical systems is the bifurcation analysis. Numerous investigations have
employed the method of bifurcation diagrams to study the effect of different
control parameters on the bubble behavior. These studies, however, focused mainly
on investigating the subharmonic (SH) and chaotic oscillations of the bubbles.
Super-harmonic (SuH) and ultra-harmonic (UH) bubble oscillations remain
under-investigated. One reason is that the conventional method used for
generating bifurcation diagrams cannot reliably identify features that are
responsible for the identification of SuH and UH oscillations. Additionally, the
conventional method cannot distinguish between the UHs and SHs. We introduce a
simple procedure for the generation of bifurcation diagrams to address this
shortcoming. This method selects the maxima of the bubble oscillatory response
and plots them alongside the traditional bifurcation points for the corresponding
control parameter. Through applying this method, the oscillatory behavior of the
bubble oscillator is analyzed, and stable SuH and UH bubble oscillations are
investigated. Based on this new analysis, the conditions for the generation and
amplification of UH and SuH regimes are discussed. 
\end{abstract}
}

\begin{document}

\maketitle


\section{Introduction} The acoustic bubble \cite{1,2,3,4,5,6,7,8,9,10,11,12} oscillator is present in many physical phenomena and
applications. Bubbles are involved in physical phenomena associated with
underwater acoustics and oceanography \cite{2,12}. Bubbles are used as catalysts for
chemical reactions in sonochemistry \cite{13,14,15,16} and several non-chemical cleaning
applications \cite{17}. Bubble oscialltions drive sonoluminscence \cite{15}, and are the
basis of several medical applications including, but not limited to, blood vessel
imaging and treatment monitoring \cite{18,19}, drug delivery \cite{20}, blood brain barrier opening \cite{21}, high
intensity focused ultrasound \cite{22}, shock wave lithotripsy \cite{23} and histotripsy
and clot lysis\cite{24}.

The bubble oscillator is a highly nonlinear dynamical system \cite{1,2,3,4,5,6,7,8,9,10,11,12, 25,26,27,28,29,30,31,32,33,34,35,36,37,38,39,40,41}; the
oscillatory bubble behavior has been referred to as chaotic and complex
\cite{1,2,3,4,5,6,7,8,9,10,11,12, 25,26,27,28,29,30,31,32,33,34,35,36,37,38,39,40,41}. Due to the complex behavior, a comprehehsive understanding of the
phenomena associated with bubble dynamics is difficult. Consequently, the
optimization of applications is a challenging task. Morever, due to the
incomplete knowledge on the nonlinear behavior of bubbles, many applications are
not optimized and this limits progress in the associated fields (e.g., enhanced
drug delivery \cite{20}).

Methods of nonlinear dynamics (e.g. resonance curves and bifurcation diagrams)
have been extensively applied to investigate bubble behavior \cite{1,2,3,4,5,6,7,25,26,27,28,29,30,31,32,33,34,35,36,37,38,39,40,41}. It has
been shown that the bubble oscillator can exhibit $\frac{1}{2}$,$\frac{1}{3}$,$\frac{1}{4}$,$\frac{1}{5}$ or
higher order SHs, as well as period doubling route to chaos \cite{1,2,3,4,5,6,7,25,26,27,28,29,30,31,32,33,34,35,36,37,38,39,40,41}. These
studies have shed light on the nonlinear dynamics and bifurcation structure of
the bubbles; however, the approaches used in these publications have provided
insights primarily on SH and chaotic bubble oscillations. Details of the
super-harmonic (SuH) and ultra-harmonic (UH) oscillations remain poorly
understood. One of the reasons is that the conventional analysis method only
extracts the data after every period of acoustic driving pressure \cite{1,5}. Analysis
methods need to be developed to identify and explore SuH and UH oscillations
alongside the SH and chaotic regimes.\\

In this work, we introduce a more comprehensive and simple method to study the
SuH and UH bubble oscillations. The bifurcation structure of the bubble
oscillator is constructed by plotting the maxima of the stable oscillations of
the bubble; which is plotted alongside the conventional bifurcation diagram.
Using this method we were able to straightforwardly identify the SuH and UH
oscillations and explore the conditions that are required to generate and amplify
the SuH and UH oscillations.

This method establishes a framework that provides a more comprehensive
understanding of the rich nonlinear behavior of bubbles; consequently, it may
help in optimizing current applications and/or can be used to discover new
nonlinear bubble behaviors that may result in new applications.
\section{Methods}
\subsection{The Bubble model}
The radial oscillations of the bubbles are numerically simulated by solving the
well known Keller-Miksis equation \cite{43}:

{\raggedright

$$\rho{}\left(1-\frac{\dot{R}}{c}\right)R\ddot{R}+\frac{3}{2}\rho{}{\dot{R}}^2\left(1-\frac{\dot{R}}{3c}\right)=$$
\begin{equation}
\left(1+\frac{\dot{R}}{c}+\frac{R}{c}\frac{d}{dt}\right)\left((p_0+\frac{2\sigma{}}{R_0}){\left(\frac{R_0}{R}\right)}^{3k}-\frac{2\sigma{}}{R}-\frac{4\mu{}\dot{R}}{R}-P_0+P_Asin\left(2\pi{}ft\right)\right)
\end{equation}

In this equation, R is radius at time t, $R_0$is the initial bubble radius,
$\rho{}$is the liquid density, c is the sound speed, $p_0$ is the atmospheric
pressure, $\sigma{}$ is the surface tension, $\mu{}$ is the liquid viscosity,
$P_A$ and \textit{f} are the amplitude and frequency of the applied acoustic
pressure.
}

{\raggedright
Oscillations of a bubble generates a backscattered pressure ($P_{Sc}$) which can
be calculated by \cite{44}:
}

\begin{equation}
 P_{Sc}=\rho\left(\frac{R}{d}\right)(R\ddot{R}+2{\dot{R}}^2)
\end{equation}

{\raggedright
where $d$ is the distance from the center of the bubble (and for simplicity is
considered as 1m in this paper) \cite{44}.
}

Equation 1 is solved using the 4$^{th}$ order Runge-Kutta technique; the control
parameters of interest are $R_0$, \textit{f }and $P_A$.  The resulting radial
bubble oscillations are visualized using the bifurcations diagrams. Bifurcation
diagrams of the normalized bubble oscillations $\frac{R}{R_0}$ are presented as
a function of driving pressure. Detailed analysis is presented at select control
parameters using a) the radius versus time curves, b) phase portrait analysis and
c) the frequency spectrum of the backscattered pressure.
\subsection{Bifurcation diagrams}
For highly nonlinear systems like bubble oscillators, small changes in the
initial conditions of the system or control parameters can result in large
changes in the behavior of the system. However, many studies focus on analyzing
the dynamics of the bubbles over discrete values of the control parameters. Due
to the complexity and sensitivity of the bubble dynamics to the exposure
parameters and initial conditions, the limited values of control parameters used
in these studies do not provide a comprehensive understanding of the bubble
dynamics. In addition, due to the discrete nature of the investigated parameters,
many exposure parameter combinations have not been investigated. Bifurcation
diagrams are valuable tools to analyze the dynamics of nonlinear systems where
the qualitative and quantitative changes of the dynamics of the system can be
investigated effectively over a wide range of the control parameters.

\subsubsection{Conventional bifurcation analysis}

When dealing with systems responding to a driving force, to generate the points
in the bifurcation diagrams vs. the control parameter, one option is to sample
from a specific point in each driving period. The approach can be summarized in:
\begin{equation}
Q \equiv (R(\Theta))\{(R(t),  \dot{R}(t) ):\Theta= \frac{n}{f} \} \hspace{0.5cm} where \hspace{0.5cm} n=100,101...150
\end{equation}
Where $Q$ denotes the points in the bifurcation diagram, $R$ and $\dot{R}$
are the time dependentradius and wall velocity of the bubble at a given
set of control parameters of ($R_{0}$, $P_{0}$, $P_{A}$, $c$, $k$, $\mu$,
$\sigma$, $f$) and $\Theta$ is given by $\frac{n}{f}$.  Points on the bifurcation
diagram are constructed by plotting the solution of $R(t)$ at time points that are
multiples of the driving acoustic period. The results are plotted for $n=100-150$
to ensure a steady state solution has been reached.

\subsubsection{Method of peaks}
As a more general method, bifurcation points can be constructed by setting one
of the phase space coordinates to zero:
\begin{equation}
  Q \equiv max(R)\{(R, \dot{R} ):\dot{R}= 0\}
\end{equation}
In this method, the steady state solution of the radial oscillations for each
control parameter is considered. The maxima of the radial peaks ($\dot{R}=0$) are
identified (determined within 100-150 cycles of the stable oscillations) and are
plotted versus the given control parameter in the bifurcation diagrams.

The bifurcation diagrams of the normalized bubble oscillations ($\frac{R}{R_0}$) are
calculated using both methods a) and b). When the two results are plotted
alongside each other, it is easier to uncover more important details about the
SuH and UH oscillations, as well as the SH and chaotic oscillations.

\section{Results}

We have considered a bubble with initial diameter of 4 microns. The linear
resonance frequency (f$_{r}$) of the bubble is \textasciitilde{}2.04 MHz. We have
studied the bifurcation structure of the bubble in the form of $\frac{R}{R_0}$  as a
function of the driving acoustic pressure for (0.2f$_{r}$$<$f$<$3f$_{r}$) and
(1kPa$<$P$_{A}$$<$3 MPa). Results are shown in Figure 1.  Figure 1a-c shows the
bifurcation structure of the bubble versus P$_{A }$ for f=0.7, 1.2 and 2.6 MHz
respectively. To focus on more practical and stable oscillation regimes, we have omitted the
parameter ranges that result in chaotic oscillations or bubble destruction
($\frac{R}{R_{0}}>2$ as discussed in detail in \cite{16}). The bifurcation structures that
are produced using the conventional method are presented in blue, and the ones
produced by the method of peaks are shown in red.

Fig. 1a shows the response of the bubble when f=0.7 MHz. The conventional method
reveals a period 1 solution for P$_{A}$$<$118 kPa and detects the generation of
period 2 solution for P$_{A}$$>$120 kPa. On the other hand, the peaks method
reveals the generation of two maxima at 24 kPa$<$P$_{A}$$<$56 kPa and three
maxima for 56 kPa$<$P$_{A}$$<$118 kPa. The three maxima undergo period doublings
(PDs) resulting in a period 6 solution for PA $>$ 118 kPa.
\begin{figure*}
\begin{center}
 \scalebox{0.2}{\includegraphics{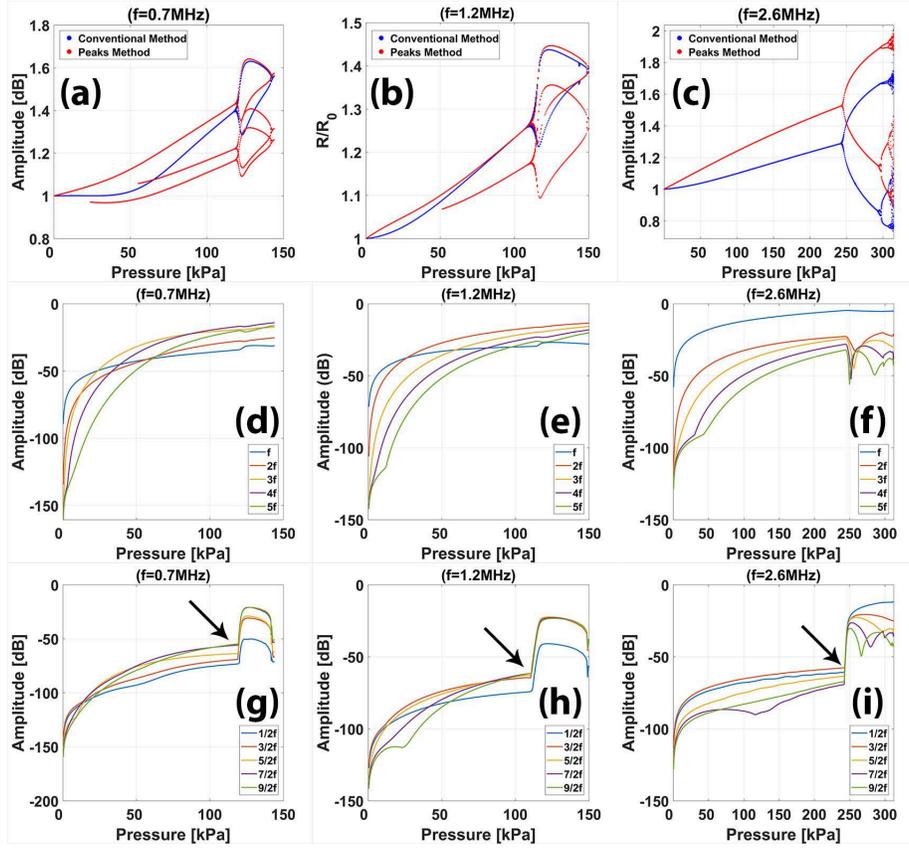}}
\caption{Bifurcation structure of the normalized radial
oscillations versus acoustic pressure of a 4 micron air bubble immersed in water
as constructed by the conventional method (blue) and the peaks method (red): a)
f=0.7 MHz, b) 1.2 MHz and c) 2.6 MHz. Harmonics of the backscattered pressure
versus acoustic pressure when d) f=0.7 MHz e) 1.2 MHz and f) 2.6 MHz. SH and UH
amplitudes of the backscattered pressure versus the acoustic pressure when g)
f=0.7 MHz h) f=1.2 MHz i)f=2.6 MHz.}
  \end{center}
\end{figure*}
Examination of the frequency content of the backscattered pressure reveals the underlying phenomenon which results in the discrepancy between the two
methods. Figure 1d shows the amplitude of the harmonics and UHs of the
backscattered signal. The occurrence of the maxima correlates with the resonance
of the harmonic contents of the signal. After a pressure threshold
(\textasciitilde{}25 kPa), the 3$^{rd}$ SuH of the backscattered signal becomes
stronger than the fundamental and other SuH harmonics, showing a 3$^{rd}$ SuH
resonance. The 3$^{rd}$ SuH saturates for $P_A$ $>$ 56 kPa concomitant with the
occurrence of 3 maxima in the peaks methods. The SH and UH contents of the
backscattered signal are shown in Fig. 1g. The simultaneous appearance of period
doublings (PDs) in the blue curve and multiple PDs in the red curve are
coincident with a sharp increase in the SH and UH content of the backscattered
signal (Fig. 1g, arrow); the backscatter at the $\frac{7}{2}$ (purple) and $\frac{9}{2}$ UHs (green)
are the strongest whit the $\frac{1}{2}$ SH (blue) the weakest component.

Fig. 1b shows that when sonication frequency is 1.2 MHz the conventional method
depicts a similar behavior to f=0.7 MHz (fig. 1a); a linear response is observed
for $P_A$ $<$ 110kPa and radial oscillations undergo a PD for $P_A$ $>$ 110 kPa. The
method of peaks reveals a solution with one maximum for PA $<$ 51 kPa which is
similar to the conventional method; above this pressure, however, 2 maxima occur
in the bifurcation diagram up until $P_A$ = 110kPa which are not detected in the
conventional method. For PA $>$ 110 kPa, the oscillations undergo two concomitant
PDs resulting in a solution with 4 maxima. The conventional method predicted the
same behavior for the two frequencies (0.7 and 1.2 MHz), however, the method of
peaks revealed more intricate details of the bubble dynamics. Fig. 1e shows that
the second harmonic of the backscattered signal has the strongest amplitude and
saturates concomitant with the generation of the initial two maxima at
\textasciitilde{}52 KPa. Fig. 1h illustrates a sharp increase in the amplitude of
SH and UHs concomitant with the generation of PDs (arrow) in the blue and red
curves as shown in Fig 1b. UH components of the signal are stronger than the SHs
(\textasciitilde{}20 dB) with $\frac{5}{2}$ and $\frac{3}{2}$ UHs being the strongest while the
$\frac{1}{2}$ SH is the weakest component.

Fig. 1c shows that when f= 2.6 MHz the conventional method (blue) and method of
peaks (red) depict a similar behavior to f=0.7 MHz and f=1.2 MHz (Fig. 1a, 1b).
The oscillations are of period 1 in both graphs until $P_A$=243 kPa; above this
pressure, PD occurs in both methods. The fundamental frequency is the strongest
frequency component in the backscattered signal as is shown in Fig. 1f. Figure 1i
indicates that the SH and UH components of the signal increase rapidly at
pressures at which the PD occurs in Fig. 1c ;  the SH component is stronger than
all the UHs.

To gain a better insight of the oscillation characteristics that the
conventional bifurcation analysis method was unable to reveal, a) radial
oscillations vs. driving acoustic periods, b) phase portraits and c)
backscattered frequency spectra are examined in detail. Exposure parameters were
chosen for which both methods give similar predictions and exposure parameters
for which the predictions diverged.
\begin{figure*}
\begin{center}
 \scalebox{0.2}{\includegraphics{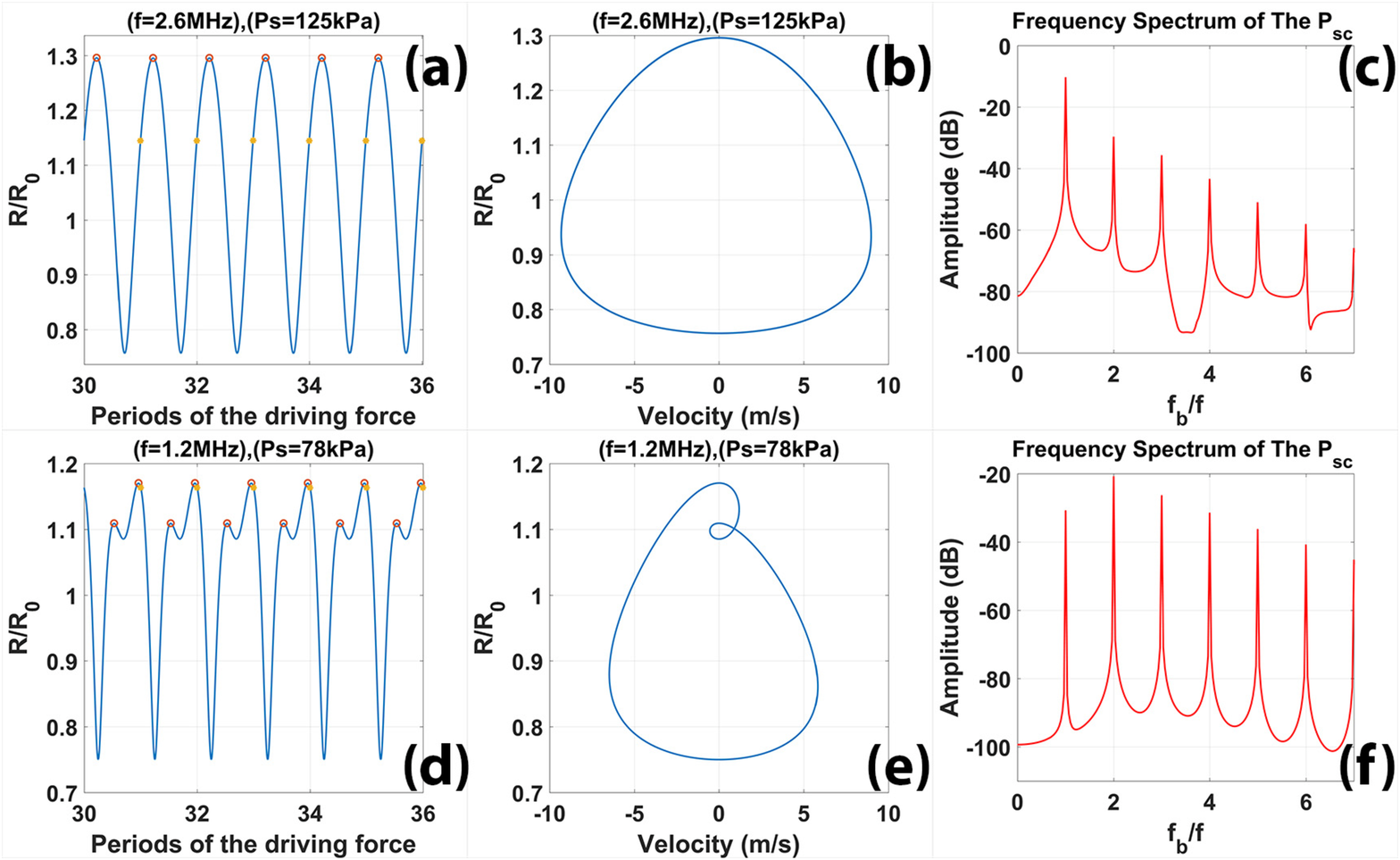}}
\caption{Oscillation characteristics of a 4 micron bubble driven at
f=2.6 MHz and 125 kPa of pressure: a) Radial oscillations versus driving periods.
Yellow dots correspond to R(t) values at each period (conventional method) and
Red dots  (peaks method) are positioned at the peaks of the R(t) curve. b) phase
portrait diagrams c) frequency spectrum of the backscattered pressure.
Oscillation characteristics of a 4 micron bubble driven with f=1.2 MHz and 78 kPa
of pressure: d) Radial oscillations versus driving periods (red shows the peaks
and yellow corresponds to R(t) at each period, e) Phase portrait f) frequency
spectrum of the backscattered pressure ($f_{b}$ is the frequency of the $P_{Sc}$).}
  \end{center}
\end{figure*}

Fig. 2a depicts the radial oscillations of the bubble for $P_{A}$=125 kPa and
f=2.6MHz. Inspection of Fig. 1c indicates that a period 1 (P1) oscillation regime is
expected. The yellow stars represent the amplitude of radial oscillations after
every period, and the red circle illustrates the maxima of the curve. There
exists only one value for all red circles and yellow stars; therefore, the signal
is a P1 signal with one maximum. The corresponding phase portrait in Fig. 2b is a
semi circular orbit, and the fundamental component of the $P_{Sc}$ is stronger than
the SuHs.\\
\begin{figure*}
\begin{center}
 \scalebox{0.2}{\includegraphics{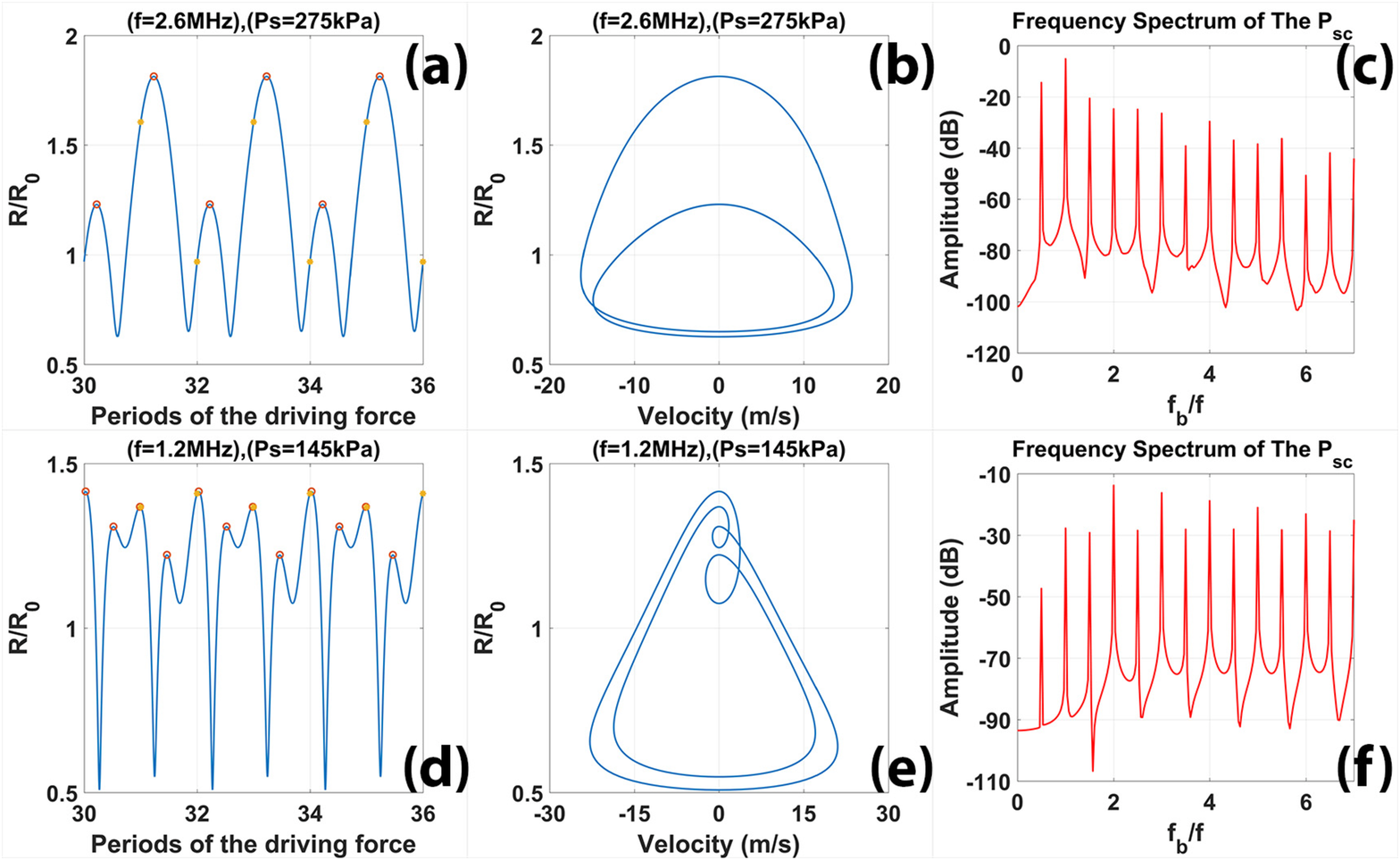}}
\caption{Oscillation characteristics of a 4 micron bubble driven
with f=2.6 MHz and 275 kPa of pressure: a) Radial oscillations versus driving
periods (yellow corresponds to R(t) at each period and red shows the peaks)  b)
Phase portrait c)frequency spectrum of the backscattered pressure. Oscillation
characteristics of a 4 micron bubble driven with f=1.2 MHz and 145 kPa of
pressure: d) Radial oscillations versus driving periods (red shows the peaks and
yellow corresponds to R(t) at each period, e) Phase portrait f)frequency spectrum
of the backscattered pressure ($f_{b}$ is the frequency of the $P_{Sc}$).}
  \end{center}
\end{figure*}
Fig. 2d shows the R-T curve that corresponds to $P_{A}$=78 kPa and f=1.2
MHz. The signal has two maxima, while the amplitude of the signal at each driving
period remains the same. In other words, this suggests a P1 signal with two
maxima. Fig. 2e shows that the phase portrait of the bubble undergoes an internal bend;
depicting a similar behavior when SHs are present in the curve. However, the
absence of SHs are evident in the frequency spectra of the corresponding $P_{Sc}$
(Fig. 2f) while the 2$^{nd}$ harmonic has the highest value (2$^{nd}$ harmonic
resonance). In this case examination of the maxima provides more complete
information about the oscillation characteristics indicating a 2$^{nd}$ SuH
resonance.

Figure 3 compares the characteristics of two P2 oscillations; one has two maxima
while the other has 4 maxima. Fig 3a shows the R-T curve of the bubble for f=2.6
MHz and $P_A$=275kPa. The signal has two maxima (two red dots), and the signal repeats its pattern
once every two acoustic driving periods (two yellow stars). The phase portrait (Fig 3b) consists of
two circular orbits with one creating another and enclosing it within itself.
Fig. 3c depicts the existence of $\frac{1}{2}$ order SH which is stronger than
the UH components. Fig. 3d shows the R-T curve of the bubble for f=1.2 MHz and
$P_A$=145kPa; the signal is of P2 but with 4 distinct maxima. The radial
oscillations repeat their pattern once every two acoustic periods (two yellow stars), and each
pattern has 4 maxima (4 red dots). The phase portrait has two circular orbits similar to Fig
3b; however, each of these circular orbits underwent an internal bend. The
frequency content of the $P_{Sc}$ in Fig. 3f has $\frac{1}{2}$ order SHs as well as
UHs; the 2$^{nd}$ order SuH is the strongest signal, and $\frac{5}{2}$ and $\frac{7}{2}$ UHs are
stronger than $\frac{1}{2}$ order SH and other UHs.

\section{Summary and Conclusion}

We have shown that the conventional method of generating bifurcation diagrams
cannot reveal the hidden details of the oscillations that typically result in UH
and SuH resonance. We have introduced a simple alternative method to generate the
bifurcation diagrams; the method extracts the peaks of the oscillations and plots
it as a function of the given control parameter. When this method is applied
alongside the conventional method one can reveal hidden details of the bubble
oscillations and identify the parameter ranges where SuH, UH or SH oscillations
occur. We can briefly categorize the scenarios shown in this paper as follows:

\begin{enumerate}
	\item The conventional method depicts a P1 oscillation regime, and the maxima method
only reveals one maximum. In this case, the oscillation has a P1 resonance and
the fundamental frequency component is the strongest in the backscattered signal.
$\frac{1}{2}$  order SH and UHs are generated concomitant with PD in both graphs
(constructed by conventional method and method of maxima) and $\frac{1}{2}$
order SHs, or $\frac{3}{2}$  UHs are stronger than other UHs.
	\item The conventional method depicts a P1 oscillation regime, but maxima method
reveals n=2,3,\ldots{} maxima. In this case, the n-th order SuH frequency
component is the strongest in the backscattered signal. Generation of PD in the
conventional method is concomitant with the generation of n-PDs in the diagram
constructed by the peaks method; this correlates with an UH resonance behavior of
$\frac{(2n-1)}{2}$ or $\frac{(2n+1)}{2}$; in other words, these UHs are stronger than the
$\frac{1}{2}$  order SHs and other UHs.
	\item The conventional method depicts a P2 oscillation regime, but the maxima method
reveals only 1 maximum; in this case, we have a P2 resonance; $\frac{1}{2}$
order SH frequency component is the stronger than UHs.
	\item UH and SHs only exist when the conventional method predicts a P2 oscillations;
however, the method of maxima needs to be applied alongside of traditional method
to determine whether SH or UH resonance are present, as well as the order of UHs.
\end{enumerate}

The nonlinear behavior of the lipid shell enhances the generation of SHs ($\frac{1}{2}$ ,
$\frac{1}{3}$ , $\frac{2}{3}$ , $\frac{1}{4}$, $\frac{3}{4}$   \ldots{}) at very low acoustic pressures and over a
more extensive frequency range [45,46]. The behavior of lipid shell MBs are more
complex due to the nonlinear response of the encapsulating shell. Implementation
of the proposed method in this paper can shine a brighter light on the behavior of
lipid coated MBs. Consequently, this approach can be used to optimize the wide
range of applications that employ coated MBs. For example, the techniques
presented can be used to optimize contrast enhanced techniques employing super
harmonics \cite{47,48} or ultraharmonics \cite{49}.

Detailed studies on the effect of initial conditions (ICs) (R(0) and $\dot{R}(0)$) on
the dynamical evolution of the bubble oscillations \cite{28,29,30,31,32,33,34} have resulted in the
discovery of new nonlinear features \cite{29,30,31,32,33,34}. For example, depending on the ICs,
the bubble has shown to exhibit period 1 (P1), P2 or P3 oscillations \cite{31,33}. Application of
the method proposed in this paper can help to better understand and categorize
these nonlinear features, especially in the SuH and UH oscillation regimes.
Results using this approach may be used to optimize applications by sending the
proper pre-conditioning pulses to manipulate the ICs of the bubble to achieve the
desirable behavior.


\end{document}